\documentclass[submission]{sacjsub}
\usepackage{amssymb}
\usepackage{hyperref}
\usepackage{xcolor}
\usepackage{subcaption}

\begin{document}

\SACJtitle{A method of generating bespoke optimised keyboard layouts that significantly reduce typing effort for Twitter users}

\SACJauthor[r]{E. Elson}{elson.e.c@gmail.com}{corresponding}

%

\SACJaddress{Department of Physics \& Astronomy, University of the Western Cape, Robert Sobukwe Rd, Bellville, Cape Town}
\SACJrunningheader{E. Elson}{Optimised keyboard layouts for Twitter users}

\SACJcitationauthors{E. Elson (eelson@uwc.ac.za)}

\SACJabstract{This study addresses the problem of generating an optimised keyboard layout for single-finger typing on a smartphone. It offers Twitter users a tweet-typing experience that requires less effort and time. Bodies of tweet text for 85 popular Twitter users are used. While existing studies have produced optimisations that may generally benefit a variety of users, this study is unique in the sense that a bespoke optimised keyboard layout is generated for each Twitter user based on their own tweets, thereby uniquely benefiting them more than other users. The optimisation process is based on moving only six letter keys from their positions on the QWERTY keyboard, and therefore strikes an effective balance between the typing efficiency improvements offered by an optimised keyboard and the effort required to learn to use it. It is shown that a Twitter user will enjoy a reduction in typing effort of at least 13.4\%. The typical user will benefit from a 15.8\% reduction, while the highest typing effort reduction is nearly 25\%. The method presented in this study could therefore be used in practical ways to offer any Twitter user a uniquely-improved tweeting experience.}

\SACJACMCategory{Human-centred computing}{Ubiquitous and mobile computing}{h}
\SACJkeywords{interaction design, design evaluation}

\SACJmaketitle

\section{Introduction}\label{sec:Intro}
QWERTY is the dominant keyboard layout for Latin-script alphabets.  Originally created for use with typewriters nearly 150 years ago, it is still used ubiquitously for a large variety of modern day devices, perhaps most notably computers and smartphones.  For better or worse, it has become the de facto modern day layout that is used for many languages and by many people.   While almost any keyboard layout, including QWERTY, can be used effectively when typing with several fingers (perhaps up to 10 on a full-size computer keyboard), the effort associated with single-finger typing depends significantly on keyboard layout.  

The smartphone has become a very popular device used by people to communicate and interact with one another, often via a preferred social media/networking service such as Facebook, Twitter, LinkedIn, etc. Given the size of the soft keyboard of a typical smartphone, and given the fact that users often type on their smart device while executing other tasks  (such as walking or eating or commuting), many smartphone users have a preference for single-finger typing.  Hoober~(2013)\footnote{\href{https://www.uxmatters.com/mt/archives/2013/02/how-do-users-really-hold-mobile-devices.php}{https://www.uxmatters.com/mt/archives/2013/02/how-do-users-really-hold-mobile-devices.php}} showed that 49\% of people using smartphones while commuting, walking, or in waiting situations use their devices with one hand, while 36\% use one hand for support and the other for typing (with a single finger).  

Using a single finger (most typically a thumb or forefinger) to navigate a QWERTY keyboard is highly non-optimal.  The QWERTY layout has many keys making up frequently-used letter pairs being located at the extremities of the keyboard (e.g., letters e, r, t) and also far apart from one another, thereby requiring a user's finger to traverse many long key-to-key paths when typing.  Additionally, several infrequently-used letters (e.g., z, x, v, b) occupy the most accessible portions of the QWERTY keyboard.   

Several authors have addressed the challenge of generating an optimised keyboard layout that is better suited (than QWERTY) to single-finger typing.  \citet{DellAmico_2009} originally presented the problem of keyboard layout optimisation in the context of single-finger typing as being a variant of the classical quadratic assignment problem.  \citet{Bi_2010} explored a ``Quasi-QWERTY optimisation" approach based on allowing letters to move at most one key from their QWERTY position, and demonstrated a gain in typing efficiency equal to approximately half of what can be obtained by unconstrained optimisation. \citet{Bi_2012} extended the work of \citet{Bi_2010} to consider multilingual (French, German, Spanish, English, Chinese) keyboards, and showed all of them to significantly reduce text input time as compared to the QWERTY keyboard, albeit for experienced users.  \citet{Hosny_2014} combined genetic and simulated annealing algorithms to generate optimal layouts for the Arabic keyboard for applications that predominantly use a single finger.  Those investigators showed their optimised layouts to outperform other layouts in terms of typing speed.

This study addresses the single-finger-typing keyboard optimisation problem in several  ways that differ to the more traditional methods seen in the literature.  Rather than attempting to generate a single optimised keyboard layout that may benefit a broad variety of users, bespoke keyboard layouts are generated for individual users.  The users themselves are Twitter users, specifically.  Given that Twitter users are very likely to tweet using a single finger whilst carrying out other activities, they serve as a unique demographic of smartphone users who will benefit from an optimised keyboard layout.  While most existing studies have used meta-heuristic algorithms to generate optimised keyboard layouts, this study uses a deterministic approach based on combinatorics, thereby guaranteeing the keyboard layouts it generates to be the most optimal according to the imposed constraints.  Furthermore, this study aims to strike an effective balance between the typing efficiency improvements offered by an optimised keyboard and the similarity of the keyboard's layout to that of the QWERTY keyboard, and hence  the effort required to learn to use the non-QWERTY keyboard.  To this end, only 6 of the 26 letter keys are moved from their QWERTY positions.  For a sample of 85 Twitter users, it is shown that all of them benefit from a significant reduction in typing effort when using their bespoke keyboards.   The method presented in this study could therefore be used in very practical ways, perhaps by means of a stand-alone app or an add-on to the Twitter app, to offer all Twitter users an improved tweeting experience. 

The layout of this paper is as follows.  Section~\ref{sec:Method} presents the keyboard model and the method of measuring typing effort, as well the deterministic method used to generate optimised keyboard layouts.  The details of the primary data sets used in this study are given in Section~\ref{sec:Data}.  Results are gathered in Section~\ref{sec:Results}, firstly collectively for all of the Twitters users considered, and then for four individual users.  The study is summarised in Section~\ref{sec:Summary}.

\section{Methods}\label{sec:Method}
\subsection{Measurement of distance}
\begin{figure}
\begin{center}
\includegraphics[width=0.8\linewidth]{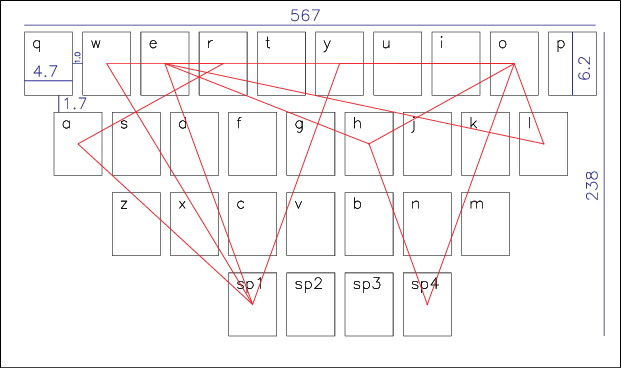}
\caption{Model of the iPhone SE (2020) English keyboard (when the screen is in portrait orientation) used in this work to measure finger-traversed distances for various samples of user-specific text.  Each of the 26 alphabet keys has a physical width and height of of 4.76~mm and 6.26 mm, respectively.  The horizontal and vertical spacings between keys are 1.01~mm and 1.70~mm, respectively.  The keyboard's width from the left edge of the Q hey to the right edge of the P key is 567~mm.  The height from the top edge of the P key to the bottom of the spacebar is 238 mm.  The spacebar has been split into 4 keys.  Shown as lines are the key-to-key paths traversed by a user's finger in order to type the sentence "Hello.  How are you?"}
\label{iphone_kb}
\end{center}
\end{figure}

This study is focused on the scenario in which a Twitter user types (i.e., tweets) using a single finger.  The effort required to type a particular word is assumed to be determined by the distance their finger traverses as it moves between keys making up various key pairs.  In order to measure the total traversed distance for a given word or body of text, a model for the built-in (standard) English keyboard for the iPhone SE (2020) is used.  Figure~\ref{iphone_kb} shows the model, while its corresponding physical dimensions are specified in the caption.  The model considers only the 26 alphabet keys and the spacebar (which is split into 4 keys). 

Given a word, or a set of words, the lengths of the straight lines joining the centres of the keys that need to be pressed are used to measure to the total traversed distance. Punctuation marks are ignored, but spaces are not.  As mentioned above, the space bar is divided into four keys.  Each time a space needs to be typed, the spacebar key closest to the key of the letter preceding the space (in the text) is used. No consideration is made for the case of a given letter, all letters are treated as being lower case.  As a simple graphical demonstration of the above-mentioned procedure, Figure~\ref{iphone_kb} shows the various key-to-key straight-line paths making up the sentence "Hello.  How are you?".  The total traversed distance for this sentence is 32.1~cm.  The sequence of key transitions is h$\rightarrow$~e$\rightarrow$~l\footnote{Two of the same letter adjacent to one another (i.e., ``ll'' from the word ``Hello'' in this example) results in a zero contribution to the total traversed distance.}$\rightarrow$~o$\rightarrow$~sp4$\rightarrow$~h$\rightarrow$~o$\rightarrow$~w$\rightarrow$~sp1$\rightarrow$~a$\rightarrow$~r$\rightarrow$~e$\rightarrow$~sp1$\rightarrow$~y$\rightarrow$~o$\rightarrow$~u.  Given that only alphabet characters are considered for a given body of text, the total traversed distances presented in this study serve as lower limits of the true corresponding values.

\citet{Fitts_1954} showed the effort of typing two symbols to be 
\begin{equation}
\alpha + \beta\log_2\left({D\over A} +1\right), 
\end{equation}
where $D$ is the distance between the keys the two symbols are assigned to and $A$ is the area of the second key.  $\alpha$ and $\beta$ are empirically determined constants, reflecting the performance of the input system tested.  The literature presents a large variety of values for $\alpha$ and $\beta$ - there is no proper consensus on what the best set of values should be. Given that the model keyboard used in this study consists of keys that all have the same area,  the key-to-key distance ($D$) is  used as a measure of the effort of typing two symbols.  The total typing effort for a given body of text is the sum of all the key-to-key distances. 

\subsection{Determining optimal layouts}
Given a keyboard model and and a method of measuring typing effort, it is possible to compare various keyboard layouts generated by repositioning a subset of the keys of a QWERTY keyboard.  The main aim of this study is to demonstrate the effectiveness of a new method of determining the optimal keyboard layout for individual, real-world Twitter users, and to quantify the reduction in typing effort afforded by a user's optimal keyboard layout.  

Keyboard optimisation studies usually adopt stochastic/heuristic methods of determining optimal layouts (e.g., \citealt{Hosny_2014}).  While such approaches do offer various benefits (such as being able to consider models based on a very large parameter space), there are also significant drawbacks.  One particular drawback is the strong possibility that the stochastic method yields a keyboard layout that is indeed improved relative to some initial layout, yet which is not the absolute best layout that offers the best level of typing performance (according to a particular objective function used to quantify typing effort).  In other words, the solution it offers represents a local minimum in the parameter space being investigated, not the global minimum. 

In this study, a deterministic approach to finding an optimal keyboard layout is adopted.  The approach is afforded by an underlying analysis goal: to offer a particular Twitter user a keyboard layout that is tailored to their unique tweet-typing style in a way that significantly boosts their tweet-typing performance, yet which also offers a high degree of familiarity in terms of the de facto QWERTY layout.  To this end, the method considers each of the 2600 combinations of three keys that can be chosen from the 26 keys of the alphabet letters, and swaps each of the three keys with those of the 2599 remaining key triplets.  Therefore, for a given body of text for a particular Twitter user, the method evaluates the typing effort for all $2600\times 2599 = 6~757~400$ keyboard layouts that can be generated by swapping three of the 26 alphabet keys with three others. This deterministic method is guaranteed to find the absolute best solution that can be generated by moving a total of only six keys.  In Section~\ref{sec:Results} it will be shown that this approach, which serves to only slightly adjust the QWERTY layout, can lead to relative improvements in tweet-typing effort of up to nearly 25\% for individual users.

\section{Data}\label{sec:Data}
Appropriate data sets are required to carry out meaningful tests of any keyboard optimisation approach.  While most keyboard optimisation studies typically use text extracted from highly formal written works (e.g., classic novels or Wikipedia) to test the performance of optimised keyboard layouts, this study uses the data of  Twitter users.  It is assumed that most Twitter users will have a preference for  single-finger typing, especially when carrying out other activities at the same time.

The 100 most recent tweets of the top 100\footnote{In terms of numbers of followers.} Twitter users as listed on the website \\\href{https://socialblade.com/twitter/top/100}{https://socialblade.com/twitter/top/100} at the time of carrying out this study were downloaded.  The accounts of some of the these users were blocked by Twitter, and hence unavailable.  Users were eliminated if a large fraction of their Tweets consisted of letters that were not from the Roman alphabet.  For those with a small fraction of tweets written with letters from other alphabets, the subset of tweets written in Roman alphabet letters was used.  All of a user's retweets were ignored, as well as all URL references to web sources.   The tweets from a final subset of 85 of the top 100 users were used in this study.  The large majority of the tweets from these users were written in English, however some tweets were written in other languages such as Spanish and Portuguese.  While most of the 85 users considered in this study are individuals, some are organisations.  A full list of the Twitter users whose data was used in this study is presented in Table~\ref{twitter_users}.  

\begin{table}
\caption{List of 85 Twitter users for whom bespoke optimal keyboard layouts were generated using their tweets data.}
\begin{center}
\begin{tabular}{lllllllllll}
BarackObama	&	britneyspears	&	espn	&	ManUtd	&	deepikapadukone	\\
katyperry	&	nytimes	&	NiallOfficial	&	KendallJenner	&	TheEconomist	\\
rihanna	&	ddlovato	&	SportsCenter	&	iHrithik	&	PlayStation	\\
elonmusk	&	shakira	&	KylieJenner	&	HillaryClinton	&	kourtneykardash	\\
Cristiano	&	KingJames	&	Drake	&	Pink                                                                                    	&	Google	\\
ladygaga	&	jimmyfallon	&	NBA	&	onedirection	&	ShawnMendes	\\
narendramodi	&	BBCBreaking	&	Harry\_Styles	&	zaynmalik	&	BIGHIT\_MUSIC	\\
YouTube	&	imVkohli	&	BBCWorld	&	kanyewest	&	M10	\\
KimKardashian	&	SrBachchan	&	sachin\_rt	&	NFL	&	NICKIMINAJ	\\
selenagomez	&	MileyCyrus	&	KevinHart4real	&	khloekardashian	&	whindersson	\\
cnnbrk	&	JLo                                                                                     	&	wizkhalifa	&	aliciakeys	&	andresiniesta8	\\
jtimberlake	&	akshaykumar	&	Louis\_Tomlinson	&	KAKA	&	Reuters	\\
Twitter	&	realmadrid	&	JoeBiden	&	ConanOBrien	&	ActuallyNPH	\\
NASA	&	BeingSalmanKhan	&	premierleague	&	EmmaWatson	&	WhiteHouse45	\\
BillGates	&	Oprah	&	LilTunechi	&	NatGeo	&	POTUS	\\
CNN	&	BrunoMars	&	LiamPayne	&	priyankachopra	&	pitbull	\\
neymarjr	&	iamsrk	&	instagram	&	Adele	&	SpaceX 	\\
\end{tabular}
\end{center}
\label{twitter_users}
\end{table}%

For each user, the first 1200 characters of their download tweets were used to generate an optimal keyboard layout.  However, given that only alphabet letters were used to quantify typing effort (i.e., grammar, numbers and emojis were ignored), the final set of useable letters was different for each user.  While the large majority of users have more than 1000 useable letters that were used to determine their optimal keyboard layout, some have only a few hundred.  In Section~\ref{sec:Results} it is shown that the reduction in tweet-typing effort offered by a user's optimised keyboard is not correlated with the number of usable letters used to generate the keyboard.

\section{Results}\label{sec:Results}
\subsection{All users}
\begin{figure}[!h]
\begin{center}
\includegraphics[width=\linewidth]{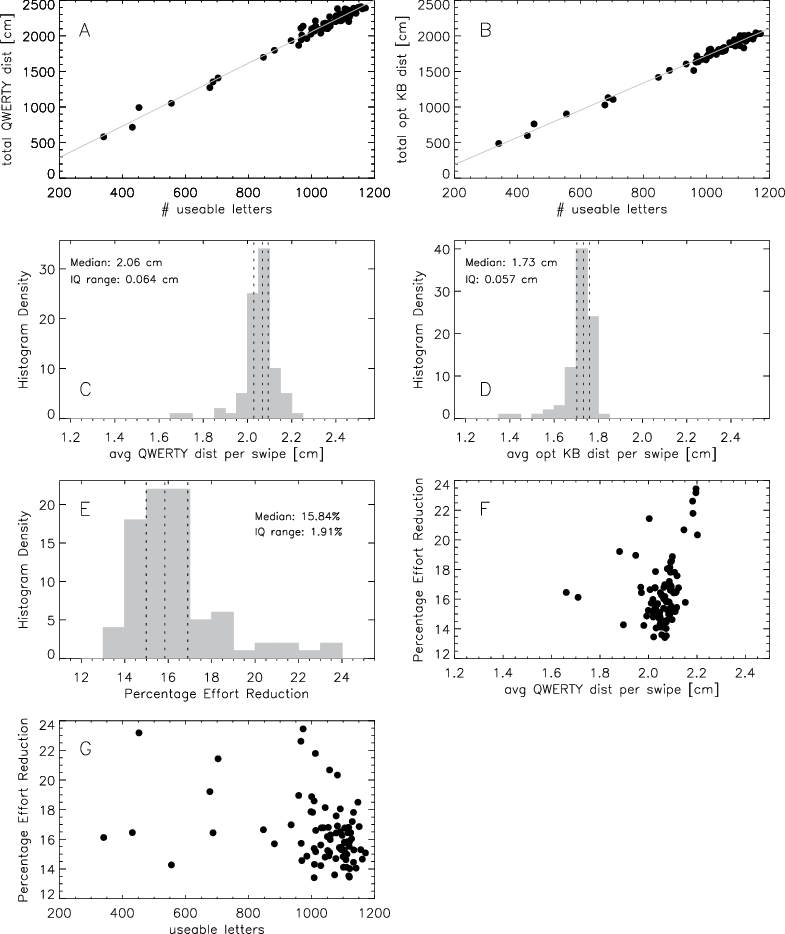}
\caption{Keyboard optimisation results for all of the 85 Twitter users considered in this study.   Panels A and B: total traversed distance as a function of number of useable characters (i.e., alphabet letters) in a Twitter user's tweet text, for QWERTY and optimised keyboards.  Panels C and D: distributions of average key-to-key distances for QWERTY and optimised keyboards.  Panel E: distribution of Percentage Effort Reduction ($PER$, defined by Equation~\ref{eqn_PER}) of users offered by their optimised keyboards.  Panel F: $PER$ as a function of average key-to-key distance between keys on a QWERTY keyboard.  Panel G: $PER$ as a function of number of useable characters in  a Twitter user's tweet text.}
\label{all_users_results}
\end{center}
\end{figure}

This section presents and discusses the tweet-typing effort reductions of all of  the 85 users considered in this study, as functions of different quantities.   Various graphical representations of the results are shown in Figure~\ref{all_users_results}.  For each user, the typing efforts related to two keyboards are considered: 1) the standard QWERTY keyboard as shown in Figure~\ref{iphone_kb}, and 2) their best optimised keyboard that results from swapping three pairs of keys (from their QWERTY positions). 

Panels A and B of Figure~\ref{all_users_results} show the total traversed distance as a function of the number of useable letters in a user's tweet text set, for QWERTY and optimised keyboards, respectively.  In both cases, the total traversed distance increases linearly with the number of  letters, as expected.  However, the slopes of the linear relations differ significantly.  In each of panels A and B, the solid line represents a first-order polynomial fit to the data: $y=mx+c$.  For the QWERTY keyboard $m=2.21$, while for the optimised keyboards $m=1.92$.  Thus, on average, the rate at which total tweet-typing effort increases with number of letters is a factor $2.21/1.92=1.15$ greater for the QWERTY keyboard than it is for a user's optimised keyboard. 

In the context of tweets that are limited to 280 characters, the above-mentioned result is perhaps not particularly relevant.  Panels C and D of Figure~\ref{all_users_results} therefore show distributions of the average traversed distance per key-to-key movement for the QWERTY and optimised keyboard cases, respectively.  Very noticeable are three important facts: 
\begin{enumerate}
\item Neither distribution is symmetric.  They each have a leftward-extending tail. 
\item The median distance for the QWERTY keyboard (2.06~cm) is larger than the median distance for the optimised keyboards (1.73~cm).  Thus for a given tweet, the typical user will incur a typing effort that is a factor $2.06/1.73 = 1.19$ greater for a QWERTY keyboard than it would be for their optimised keyboard. 
\item The distribution of QWERTY keyboard average key-to-key distances has a larger spread than the distribution of optimised keyboard distances.  The interquartile ranges of the two distributions are 0.064~cm and 0.057~cm.  Thus, users typically achieve closer-to-optimal typing performances more consistently with an optimised keyboard than they do with a QWERTY keyboard. 
\end{enumerate}

Panel E of Figure~\ref{all_users_results} shows the distribution of tweet-typing effort improvement (i.e., reduction) for the 85 users considered in this study.  The absolute reduction in tweet-typing effort offered by a user's optimised keyboard is normalised by the absolute effort for a QWERTY keyboard.  This quantity is called Percentage Effort Reduction ($PER$):
\begin{equation}
PER=100\times {D_\mathrm{qwerty} - D_\mathrm{opt}\over D_\mathrm{qwerty}},
\label{eqn_PER}
\end{equation}
where $D_\mathrm{qwerty}$ and $D_\mathrm{opt}$ are the average key-to-key distances traversed by a user for the QWERTY keyboard  and their optimised keyboard, respectively.  In other words, $PER$ is a measure of the percentage difference in typing effort between the two keyboards relative to the typing effort of the QWERTY keyboard.  Some statistics for $PER$ are as follows: 1) The minimum $PER$ is 13.4\%, while the maximum is 23.4\%, 2) The median $PER$ is 15.8\% while the interquartile range of the set of 85 $PERs$ is 1.91\%.  

These results are significant.  They demonstrate that a Twitter user's tweet-typing effort can be reduced by at least 13.4\%, and up to nearly 25\%, by moving only six keys of the traditional QWERTY keyboard. A typical user will enjoy a 15.8\% reduction in typing effort. While other studies of optimal keyboard layouts typically test their methods on text samples taken from novels or other highly formal works of literature, this study presents a method of significantly reducing the tweet-typing efforts of real-world Twitter users while still offering  77\% (i.e., $100\times20/26$) of the familiarity of a QWERTY keyboard layout.  The learning curve is therefore low, yet the reduction in tweet-typing effort high.

Panel F of Figure~\ref{all_users_results} shows  $PER$ as a function of  average key-to-key distance.  While the large majority of users fall within a small $PER$ range, the existence of a positive correlation between $PER$ and key-to-key distance is clear.  This means that, on average, the reduction in tweet-typing effort a Twitter user will enjoy by using their optimised keyboard will actually increase with the degree to which a QWERTY keyboard inhibits their typing efficiency.  Put differently: Twitter users who experience the most typing effort on a QWERTY keyboard will benefit the most from their optimised keyboard (as generated according to the method presented in this study).

Because the number of useable letters from the tweets of this study's Twitter users is non-constant, a possibility exists for the results to depend systematically on the number of useable letters.  However, Panel~G of Figure~\ref{all_users_results} shows this not to be the case.  When considering Percentage Effort Reduction as a function of the number of useable letters, there is no evidence of a correlation between the two quantities. Thus, the effectiveness of the keyboard optimisation method presented in this study is not dependent on the size of the body of text used to calculate an optimised keyboard layout. 

Finally, given that the definition of typing effort used in this study is based on the distance traversed by a Twitter user's finger between keys making up key pairs, the typing effort should be proportional to the amount of time required to generate a tweet.  Hence, by using an optimised keyboard layout based on the unique style and content of their tweets, a Twitter user could reduce the amount of time required to generate a tweet by up to nearly 25\%, which could potentially result in them producing a greater total number of tweets within a given interval of time (e.g., within a day).

\subsection{Individual users}
Having generally discussed the results for all users collectively in the previous section, this section focuses on the results from four individual users, for the sake of gaining specific insights into the ways in which their optimised keyboard layouts reduce tweet-typing effort. The full range of Percentage Effort Reductions ($PER$, Figure~\ref{all_users_results}E) was split into three equal intervals delimited by the values 13.4\%, 16.7\%, 20.0\% and 23.4\%.  For each of these four delimiting values, the user with the closest-matching $PER$ was chosen.  The users are  LilTunechi, ladygaga, shakira and neymarjr, with respective $PER$ quantities of 13.4\%, 16.8\%, 20.1\% and 23.4\%.  The key swaps made to generate each user's optimised keyboard are as follows: LilTunechi  eoa$\rightarrow$~jvb\footnote{More specifically: keys e and j were swapped, keys o and v were swapped, and keys a and b were swapped.}, ladygaga  oas~$\rightarrow$~ jvb, shakira  eoa~$\rightarrow$~ fgv, neymarjr  eoa~$\rightarrow$~ ghv.

Given the way in which tweet-typing effort is defined in this study, it is prudent to consider the statistics of a user's usage of key pairs in order to better understand how they benefit from an optimised keyboard layout.  Given 26 alphabet letters and a spacebar, there are 729 possible key pairs. Tables~\ref{tu1} to \ref{tu4} present  various details and statistics related to the 15 most-used key pairs of the four users considered in this section.  In each table,  column 1 gives the  key pairs ranked in order of percentage usage.  Column 2 gives the percentage usages of the key pairs.  Columns 3 and 4 give the physical separations (in cm) between the individual keys making up the key pairs on a QWERTY keyboard and on a user's optimised keyboard, respectively.  Column 5 gives the ratios of the two physical separations from the previous two columns.  Thus, values in column 5 greater than unity represent reductions in tweet-typing efforts for particular key pairs on an optimised keyboard, as compared to a QWERTY keyboard.  The last row in each of tables~\ref{tu1} to \ref{tu4} offers two more useful statistics: the number of unique alphabet letters making up a user's 15 most-used key pairs ($N_\mathrm{let}$, in column 1), and the total percentage usage of all of a user's 15 most-used key pairs (column 2).  

Several facts are immediately apparent from the data contained in tables~\ref{tu1} to \ref{tu4}.  For all users, the key pair e-sp\footnote{sp represents the spacebar.} is used more than any other.  This is not surprising given that words based on the Roman alphabet often end with the letter e.  Thus, e will often be followed by a space.   Indeed, three of the four users have their e key swapped with another key to generate their optimised keyboard.  Their e key is moved from the top left portion of the QWERTY keyboard to a much more central location of their optimised keyboard.  This allows the e key to be used more efficiently in other key pairs, and the distance between the e key and the spacebar is reduced. Each of the four users have the majority of their 15 most-used key pairs consisting of a letter key and the spacebar (as opposed to two letter keys).  This is not surprising given that the spacebar must be pressed when transitioning from one word to another.  An optimised keyboard layout is expected to have letters contributing to frequently-used key pairs located centrally, in order to facilitate pairing with the spacebar, while infrequently-used keys are moved further out.  This modification is apparent in the optimised keyboard of all users considered in this section.  For example: the QWERTY keyboard has letters v and b located immediately above the space bar, in the centre of the keyboard.  Having these infrequently-used letters occupying prime locations on the keyboard is highly non-optimal. Indeed, for all of the users, at least one of the v and b letter keys is moved to the outer regions of the optimised keyboard, and replaced with letters making up frequently-used key pairs (letters such as o, a, s) that are located at the extremities of the QWERTY keyboard. 

\begin{table}[!h]
  \begin{subtable}[t]{.5\linewidth}%
    \centering%
    \begin{tabular}{|c|c|c|c|c|}
    \hline
    pair  &  usage  &  $D_\mathrm{qwe}$  & $D_\mathrm{opt}$  & ${D_\mathrm{qwe}\over D_\mathrm{opt}}$ \\
            &  \%        &   cm                           &  cm                          &            \\
    \hline
e-sp	&	3.01	&	2.54	&	1.59	&	1.60	\\
sp-t	&	2.72	&	2.41	&	2.41	&	1.00	\\
t-sp	&	1.98	&	2.41	&	2.41	&	1.00	\\
t-h  	&	1.73	&	1.18	&	1.18	&	1.00	\\
o-sp	&	1.69	&	2.54	&	0.80	&	3.19	\\
n-sp	&	1.61	&	0.80	&	0.80	&	1.00	\\
sp-a	&	1.57	&	2.35	&	1.40	&	1.68	\\
s-sp	&	1.44	&	1.97	&	1.97	&	1.00	\\
e-r  	&	1.28	&	0.58	&	2.18	&	0.27	\\
sp-m	&	1.28	&	2.45	&	2.45	&	1.00	\\
h-e  	&	1.28	&	2.18	&	0.58	&	3.77	\\
sp-o	&	1.24	&	3.53	&	0.98	&	3.59	\\
y-sp	&	1.24	&	2.41	&	2.41	&	1.00	\\
i-n  	&	1.15	&	1.62	&	1.62	&	1.00	\\
sp-g	&	1.11	&	1.70	&	1.70	&	1.00	\\
\hline
$N_\mathrm{let}=12$	&	24.31	&		&		&		\\
   \hline
    \end{tabular}
    \caption{LilTunechi}\label{tu1}
  \end{subtable}%
  \begin{subtable}[t]{.5\linewidth}
    \centering
    \begin{tabular}{|c|c|c|c|c|}
    \hline
  pair  &  usage  &  $D_\mathrm{qwe}$  & $D_\mathrm{opt}$  & ${D_\mathrm{qwe}\over D_\mathrm{opt}}$ \\
          &  \%        &   cm                           &  cm                          &            \\
    \hline
e-sp	&	2.97	&	2.54	&	2.54	&	1.00	\\
s-sp	&	2.50	&	1.97	&	0.80	&	2.47	\\
sp-t	&	2.27	&	2.41	&	2.41	&	1.00	\\
sp-a	&	2.12	&	2.35	&	0.98	&	2.39	\\
sp-s	&	1.97	&	1.97	&	1.40	&	1.40	\\
i-n	&	1.77	&	1.62	&	1.62	&	1.00	\\
sp-l	&	1.46	&	3.30	&	3.30	&	1.00	\\
h-a	&	1.46	&	2.89	&	0.98	&	2.94	\\
a-n	&	1.43	&	3.56	&	1.16	&	3.08	\\
sp-	&	1.39	&	3.13	&	3.13	&	1.00	\\
n-sp	&	1.39	&	0.80	&	0.80	&	1.00	\\
r-sp	&	1.39	&	2.41	&	2.41	&	1.00	\\
t-h	&	1.35	&	1.18	&	1.18	&	1.00	\\
o-u	&	1.35	&	1.16	&	0.85	&	1.36	\\
t-sp	&	1.31	&	2.41	&	2.41	&	1.00	\\
\hline
$N_\mathrm{let}=11$	&	26.13	&		&		&		\\
\hline
    \end{tabular}
    \caption{ladygaga}\label{tu2}
  \end{subtable}\par\bigskip
  \begin{subtable}[t]{.5\linewidth}%
    \centering%
    \begin{tabular}{|c|c|c|c|c|}
    \hline
    pair  &  usage  &  $D_\mathrm{qwe}$  & $D_\mathrm{opt}$  & ${D_\mathrm{qwe}\over D_\mathrm{opt}}$ \\
            &  \%        &   cm                           &  cm                          &            \\
    \hline
e-sp	&	3.18	&	2.54	&	1.59	&	1.60	\\
s-sp	&	2.95	&	1.97	&	1.97	&	1.00	\\
sp-t	&	2.42	&	2.41	&	2.41	&	1.00	\\
o-sp	&	1.99	&	2.54	&	1.59	&	1.59	\\
a-n	&	1.96	&	3.56	&	1.16	&	3.08	\\
n-sp	&	1.65	&	0.80	&	0.80	&	1.00	\\
o-n	&	1.53	&	1.81	&	1.40	&	1.29	\\
sp-a	&	1.46	&	2.35	&	0.98	&	2.39	\\
sp-d	&	1.46	&	1.70	&	1.70	&	1.00	\\
t-h	&	1.34	&	1.18	&	1.18	&	1.00	\\
t-o	&	1.30	&	2.31	&	0.85	&	2.72	\\
sp-c	&	1.27	&	0.80	&	0.80	&	1.00	\\
sp-m	&	1.27	&	2.45	&	2.45	&	1.00	\\
t-sp	&	1.23	&	2.41	&	2.41	&	1.00	\\
sp-e	&	1.19	&	2.54	&	1.18	&	1.59	\\
\hline
$N_\mathrm{let}=10$	&	26.19	&		&		&		\\
\hline
    \end{tabular}
    \caption{shakira}\label{tu3}
  \end{subtable}%
  \begin{subtable}[t]{.5\linewidth}
    \centering
    \begin{tabular}{|c|c|c|c|c|}
    \hline
  pair  &  usage  &  $D_\mathrm{qwe}$  & $D_\mathrm{opt}$  & ${D_\mathrm{qwe}\over D_\mathrm{opt}}$ \\
          &  \%        &   cm                           &  cm                          &            \\
    \hline
e-sp	&	3.75	&	2.54	&	1.59	&	1.60	\\
o-sp	&	3.32	&	2.54	&	1.59	&	1.59	\\
a-sp	&	3.02	&	2.35	&	0.80	&	2.96	\\
s-sp	&	2.07	&	1.97	&	1.97	&	1.00	\\
sp-a	&	2.03	&	2.35	&	0.98	&	2.39	\\
sp-t	&	1.94	&	2.41	&	2.41	&	1.00	\\
n-o 	&	1.55	&	1.81	&	0.98	&	1.84	\\
t-o 	&	1.47	&	2.31	&	1.18	&	1.96	\\
sp-d	&	1.38	&	1.70	&	1.70	&	1.00	\\
o-r 	&	1.34	&	2.89	&	1.65	&	1.75	\\
r-sp	&	1.34	&	2.41	&	2.41	&	1.00	\\
sp-n	&	1.29	&	1.91	&	1.91	&	1.00	\\
o-s 	&	1.25	&	3.84	&	2.31	&	1.66	\\
sp-s	&	1.16	&	1.97	&	1.97	&	1.00	\\
sp-p	&	1.12	&	3.97	&	3.97	&	1.59	\\
\hline
$N_\mathrm{let}=9$	&	28.02	&		&		&		\\
\hline
\end{tabular}
    \caption{neymarjr}\label{tu4}
  \end{subtable}
\caption{Detailed information and statistics for the 15 most-used key pairs of four different Twitter users.  In each sub-table, the columns are: 1) keys making up key pair, 2) percentage usage of key pair, 3/4) physical distance (in cm) between keys making up key pair on QWERTY/optimised keyboards, 5) ratio of physical distances presented in previous two columns.  In the bottom row of each table there appears the total number of unique letters ($N_\mathrm{let}$) making up a user's 15 most-used letter key pairs, and total percentage usage of all 15 most-used letter pairs.}
\label{letter_pairs_table}  
\end{table}

With the exception of a single letter pair for one of the users\footnote{Letter pair er for user LilTunechi.}, all 15 most-used key pairs  of the four users are associated with shorter key-to-key distances on their optimised keyboard than on the QWERTY keyboard.  Considering user neymarjr as an example: the top 5 key pairs have associated key-to-key distances that are factors 1.60, 1.59, 2.96, 1.00 and 2.39 greater on a QWERTY keyboard than on their optimised keyboard.  It is obviously this reduction in the lengths of frequently-used key pairs that allows an optimised keyboard layout to offer a reduced tweet-typing effort.  All users have at least 5 of their 15 most-used key pairs associated with key-to-key distances that are a factor $>1.5$ longer on the QWERTY keyboard than on their optimised keyboard.  A factor 1.5 corresponds to a 33\% reduction in the key-to-key distance on a QWERTY keyboard.  It is not surprising that the user whose optimised keyboard layout offers the highest reduction in tweet-typing effort (neymarjr, 23.4\%) has nine of their 15 most-used key pairs corresponding to a distance reduction of 33\% or more on their optimised keyboard, as compared to a QWERTY keyboard. 

Two more facts that stand out from the information presented in tables~\ref{tu1} to \ref{tu4} are as follows.   The number of unique letter keys (i.e., excluding the spacebar) that make up a user's most-used letter pairs generally decreases as tweet-typing effort is reduced.  The four users considered in this section have 12, 11, 10, 9 unique letters making up their 15 most-used letter pairs, while their overall Percentage Effort Reductions are 13.4\%, 16.8\%, 20.1\% and 23.4\%.  This result is easy to understand in the sense that given M unique letter keys that make up a significant fraction of a user's most-used letter pairs, an optimised keyboard layout based on moving N letter keys (6 in this study) will yield the best results when the difference between M and N is minimised.  Methods used to generate optimal keyboard layouts  often allow all of the 26 letter keys on a QWERTY keyboard to be moved.  However, the results from this study clearly show such an approach not to be necessary.   When the number of unique letters making up a significant fraction of a user's most-used letter pairs is small (e.g., 10), effective optimised keyboard layouts can be generated my moving only a few letter keys.  A major benefit of such an approach is the manner in which much of the QWERTY layout, and hence a user's familiarity thereof, is retained.  The second standout feature for the data in tables~\ref{tu1} to \ref{tu4} is  fact that the summed percentage usages of a user's 15 most-used key pairs increases as tweet-typing effort is reduced.  The four users considered in this section have their 15 most-used letter pairs making up 24.31\%, 26.13\%, 26.19\% and 28.02\% of all possible letter pairs.  The explanation for this is analogous to the one presented above for the correlation between the number of unique keys making up frequently-used key pairs and Percentage Effort Reduction.

\begin{figure}[!h]
\begin{center}
\includegraphics[width=\linewidth]{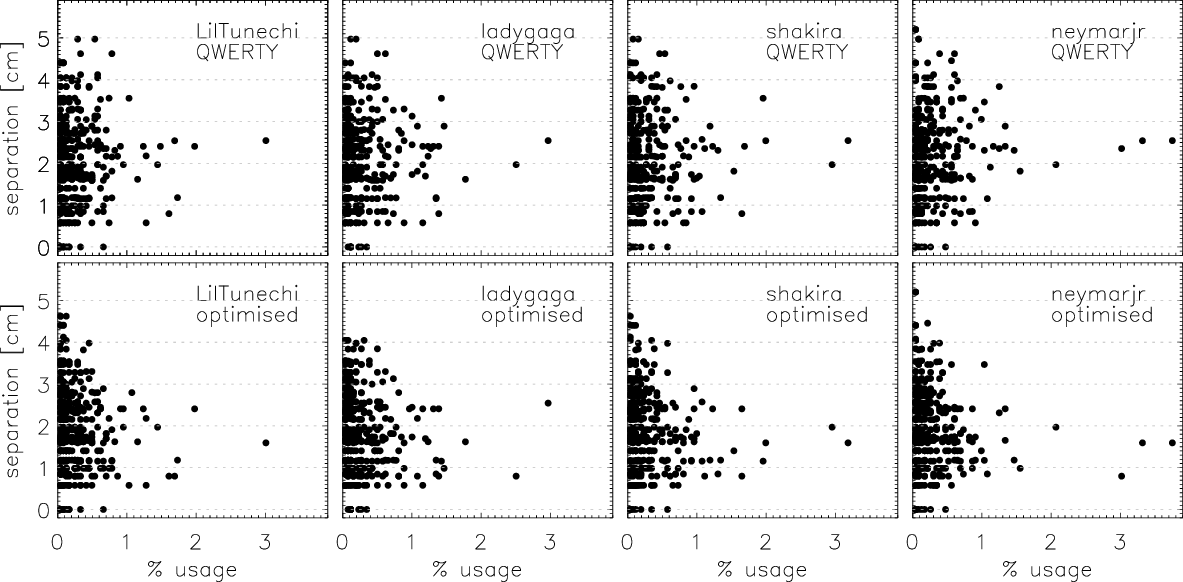}
\caption{Separation between keys making up all of a Twitter user's key pairs as a function of percentage usage of key pairs.  From left to right, columns show the results for the various users.  Top row: results for QWERTY keyboard.  Bottom row: results for optimised keyboards.  Very clear is the fact that high-usage pairs have smaller key-to-key distances on an optimised keyboard.  Keys making up very long key pairs (e.g., 5 cm) on a QWERTY keyboard and brought closer together on an optimised keyboard.}
\label{distances}
\end{center}
\end{figure}

Figure~\ref{distances} shows the physical separation (in cm) between the keys making up all of a user's key pairs as a function of the percentage usage of the pairs.  It does this for each user (left column to right column) for the QWERTY keyboard (top row) and their optimised keyboard (bottom row).   A point to note when comparing the two panels for any given user is the fact that corresponding points may shift only vertically, and not horizontally. That said, it is clear that for all users the key-to-key distances of their most frequently-used key pairs are reduced.  For example:  user neymarjr's three most-used key pairs all have key-to-key distances between 2~cm and 3~cm on a QWERTY keyboard, but less than 2 cm on the optimised keyboard.  Also, key pairs with a low percentage usage yet with a large key-to-key separation, seen in the top-left portions of the top-row panels, are notably located lower down in the bottom panels.  

\begin{figure}[!h]
\begin{center}
\includegraphics[width=\linewidth]{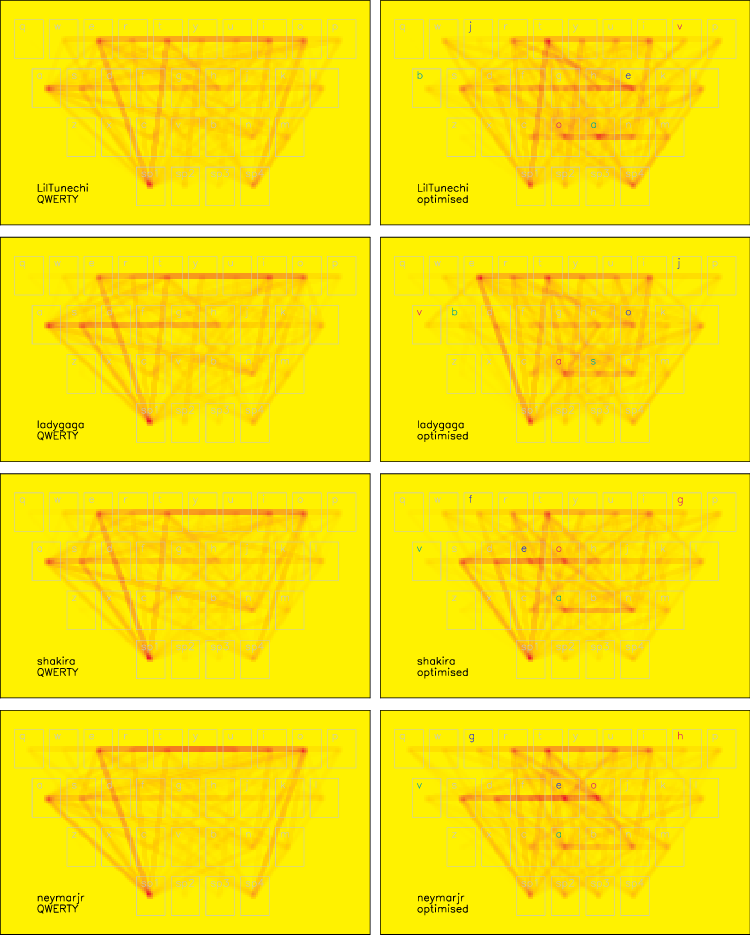}
\caption{Heat maps maps for the various Twitter users (top to bottom) showing their key-to-key movements carried out by a user's finger on QWERTY (left column) and optimised (right column) keyboards.  In all panels, the colour scheme is such that darker (i.e., redder) paths are those traversed more frequently. Each panel has the corresponding keyboard overlaid in a light-grey colour. For the optimised keyboards, the pairs of swapped letters are highlighted in the same non-grey colour (either red, green or blue).}
\label{heat_maps}
\end{center}
\end{figure}

For each user, Figure~\ref{heat_maps} shows heat maps for the QWERTY keyboard (left-side panels) and for their optimised keyboard (right-side panels).  These heat maps show the actual key-to-key movements carried out by a user's finger at they tweet.  In all panels, the colour scheme is such that darker (i.e., redder) paths are those traversed more frequently. Each panel has the corresponding keyboard overlaid in a light-grey colour. For the optimised keyboards, the pairs of swapped letters are highlighted in the same non-grey colour (either red, green or blue).  The heat maps very clearly highlight the ways in which each user's tweet-typing effort is reduced.  For all users, the QWERTY layout has a large fraction of total key-to-key movement occurring in the top key row.  This alone significantly contributes to the increased tweet-typing efforts associated with the QWERTY layout.  Also clear is the very common and long path between the e key located in the top left portion of the QWERTY keyboard and the spacebar. There are also long letter-spacebar paths that are frequently traversed for several other letter keys on the QWERTY keyboard (examples: a, t, o).  For all users, none of the letter keys in the bottom row of the QWERTY keyboard combine with the spacebar in order to form frequently-used key pairs.  Therefore, most of the paths of key pairs that have one key being the spacebar are  unnecessarily long on a QWERTY keyboard. 

All of the above-mentioned issues are address by the optimised keyboard layouts.  For all users, the amount of key-to-key movement in the top row of their optimised keyboard is significantly reduced.  In addition to minimising the need for a user to regularly extend their thumb to keys in the top row, the number of very long letter-to-spacebar movements is also effectively reduced.  Also noticeable is the fact that the key-to-key movement in the top row is  more centrally located, typically spanning 5 or 6 keys on an optimised keyboard instead of 7 on a QWERTY keyboard.  For all users, a large fraction of their total movement is shifted and limited to the middle and bottom letter rows of their optimised keyboard, making for much easier access.  Furthermore, this again significantly reduces the path lengths of key pairs that have one key being the spacebar, as well as the path lengths of various letter-letter pairs.  For all of the optimised keyboards shown in Figure~\ref{heat_maps}, it is clear that infrequently-used letter keys that are centrally located in the QWERTY layout are shifted to the extremities of the optimised keyboard, and replaced by letters that contribute to frequently-used letter pairs.  For example, v is always moved far away from its central position immediately above the spacebar on the QWERTY keyboard.  Some keys that do not contribute to frequently-used letter pairs retain their QWERTY positions in the optimised keyboards (examples: q, z, m, p).  Even though their positions make for large key-to-key path lengths for the letter pairs they contribute to, those letter pairs are rarely used.

Finally, it should be noted that, when considering all of the users from this study, those that enjoy the highest reductions in tweet-typing effort thanks to their optimised keyboard tweet mainly in languages other than English.  This important fact suggests a user's reduction in tweet-typing effort offered by their optimised keyboard may significantly depend on the language in which they tweet.  A follow-up study will aim to quantitatively investigate this possibility.

\section{Summary}\label{sec:Summary}
This study is focused on the combinatoric problem of generating an optimised keyboard layout for a given Twitter user by moving only three letter pairs from their positions on a QWERTY keyboard.  The optimisation method developed, tested and demonstrated in this study achieves this goal in a deterministic manner that guarantees optimal results.  Unlike previous studies that have focused on producing optimised keyboard layouts that are generally useful to a large variety of users,  this study's optimisation method produces a unique keyboard layout for any Twitter user, based on the content and style of their tweets.  As such, each Twitter user uniquely benefits from their optimised keyboard. By moving only six letter keys from their positions on the QWERTY keyboard, an effective balance is achieved between the reduced typing effort offered by a Twitter user's optimised keyboard and the amount of time and effort required to become accustomed to using it. The results from this study clearly demonstrate the significant reductions in typing effort offered by such an optimised keyboard.  Compared to a QWERTY layout, all of the 85 Twitter users considered in this study enjoy a tweet-typing effort reduction of at least 13.4\%.  The median reduction is 15.8\%, while the maximum is slightly less than 25\%.  The tweet-typing finger movements, and the statistics thereof, of four individual Twitter users from this study are considered and discussed.   The ways in which each user uniquely benefits from their personalised  keyboard are clearly shown, and the reasons therefore presented.   

Given that the definition of typing effort used in this study is based on the physical distance traversed by a Twitter user's finger as they tweet, the results further imply that the average amount of time required by a user to generate a tweet will be lessened, which could potentially result in them producing more tweets within a given time interval.  Thus, this study's unique optimisation method could be practically implemented in various ways in order to offer Twitter users an improved tweeting experience that reduces both effort and time.

\bibliographystyle{unsrtnat}
\bibliography{bibliography}
\end{document}